\documentclass[epj,twocolumn]{webofc}
\usepackage[varg]{txfonts}   
\wocname{???????}
\woctitle{The TimeMachineFactory}
\begin{document}
\title{On Closed Timelike Curves and Warped Brane World Models}
\author{R. J. Slagter\inst{1,2}\fnsep\thanks{\email{reinoudjan@gmail.com}}}
\institute{ASFYON, Astronomisch Fysisch Onderzoek Nederland \and Institute of Physics, University of Amsterdam, The Netherlands}
\abstract{At first glance, it seems possible to construct in general relativity theory  causality violating solutions. The most striking one is the Gott spacetime. Two cosmic strings, approaching each other with high velocity, could produce closed timelike curves. It was quickly recognized that this solution violates physical boundary conditions. The effective one particle generator becomes hyperbolic, so the center of mass is tachyonic.
On a 5-dimensional warped spacetime, it seems possible to get an elliptic generator, so no obstruction is encountered and the velocity of the center of mass of the effective particle has an overlap with the Gott region. So a CTC could, in principle, be constructed. However, from the effective 4D  field equations on the brane, which are influenced by the projection of the bulk Weyl tensor on the brane, it follows that no asymptotic conical space time is found, so no angle deficit as in the 4D counterpart model. This could also explain why we do not observe cosmic strings.}
\maketitle
\section{Introduction}\label{intro}
Physicists speculate that extra spatial dimensions could exist in addition to our ordinary 4-dimensional spacetime. The idea that spacetime could have more than four dimensions was first proposed by Kaluza and Klein (KK) in the early 20th century. These higher dimensional models can be used to address several of the shortcomings of the Standard Model, i.e., the unknown origin of dark energy and dark matter, the neutrino oscillations and the weakness of gravity (hierarchy problem). In these models, the weakness of gravity might be fundamental. One might na\"{\i}vely imagine that these extra dimensions must be very small and never observable. Recently there is growing interest in the so-called brane world models,  first proposed by  Arkani-Hamed, Dimopoulos and Dvali (ADD)\cite{ADD:1998} and extended by Randall and Sundrum (RS)\cite{RS:1999}. In these models, the extra dimension can be very large compared to the ones predicted in string theory. We live in a (3+1)-dimensional brane, embedded in a 5-dimensional bulk space time. Gravitons can then propagate into the bulk, while the other fields are confined to the branes. The weakness of gravity can be understand by the fact that it "spreads" into the extra dimension and only a part is felt in 4D. This means that all of the four forces could have similar strengths and gravity only appears weaker as a result of this geometric dilution. The huge disparity between the electro-weak scale, $M_{EW}=10^3 GeV$ and the gravitational mass scale $M_{Pl}=10^{19} GeV$ will be suppressed by the volume of the extra dimension, or the curvature in than region. The effective 4D gravitational coupling will be $M_{eff}^2=M_{EW}^{n+2}R^n$, instead of a fundamental $M_{Pl}^2$. For $n=2$ and $M_{eff}$ in order of $10^{19} GeV$, $R$ will be in order of millimeters. This effect can also be achieved in the RS model by a warp factor: $M_{eff}^2=(1-e^{-R})M_5^3$, with R the compactification  radius.
Compact objects, such as black holes and cosmic strings, could have tremendous mass in the bulk, while their warped manifestations in the brane can be consistent with observations.
With the new data from the Large Hadron Collider at CERN, it might be possible to observe these extra dimensions. An extra feature of the RS model is that it predicts electro-weakly coupled KK-modes in the TeV range.

Cosmic strings can be formed in symmetry-breaking phase transitions in the early stages of the universe\cite{vil:1994}. They consist of trapped regions of false vacuum in U(1)-gauge theories. They are topological defects, similar to flux tubes in type-II superconductors. When the temperature in the early universe decreased, the scalar field developed a locus of trapped points of false vacuum.
Density perturbations produced by these strings of GUT scale, $G\mu = \eta^2/M_{Pl}^2$, where G is Newton's constant, $M_{Pl}$ the Planck mass, $\mu$ the mass per unit length of the string and $\eta$ the symmetry breaking scale, could have served as seeds for the formation of galaxies and clusters. However, recent observation of the cosmic microwave background (CMB) radiation disfavored this scenario. The WAMP-data proves that cosmic strings cannot  contribute more than an insignificant proportion of the primordial density perturbation, $G\mu \leq 10^{-6}$. Recently, brane world scenarios suggest the existence of fundamental strings, predicted by superstring theory\cite{davis:2005,vil2:2005}. These super-massive cosmic strings, $G\mu\sim 1$, could be produced when the universe underwent phase transitions at energies much higher than the GUT scale and could play a role very similar to that of cosmic strings.
An other aspect of cosmic strings is the angle deficit they exhibit at finite distance of the core, i.e., the spacetime is there Minkowski minus a wedge\cite{garf:1985}. See figure 1. They will produce a very special pattern of lensing effect, not found yet in observations.
Because there is no z-dependency (considered in polar-coordinates), one often studies the related (2+1)-dimensional models, where the stringlike objects are treated as gravitating point particle, or "cosmons"\cite{star:1963,deser1:1983,deser2:1992}. See figure 2. These models are also studied in relation with causality problems and toy-models of quantized  gravitating point particles\cite{hooft1:2008}.
An interesting example of the richness of the (2+1)-dimensional gravity, is the Gott spacetime\cite{gott:1990}. An isolated pair of cosmic strings, equivalent with two point particles in (2+1)-dimensions, moving in opposite directions with sufficient high velocity, could generate a closed timelike curve (CTC). See figure 3. If an advanced civilization could manage to make a closed loop around this Gott pair, they will be returned to their own past.
However, it is not a surprise that one can prove that the Gott spacetime will also be present at spatial infinity, which constitutes un-physical boundary conditions\cite{deser2:1992,hooft2:1992}. Moreover, it turns out that the effective one-particle generator has a tachyonic center of mass. This means that the energy-momentum vector is spacelike.
Even a closed universe will not admit these CTC's\cite{hooft3:1993}, so the chronology protection conjecture is saved. In fact, a configuration of point particles admits a Cauchy formulation within which no CTC's are generated.
\begin{figure}
  \includegraphics[height=.25\textheight]{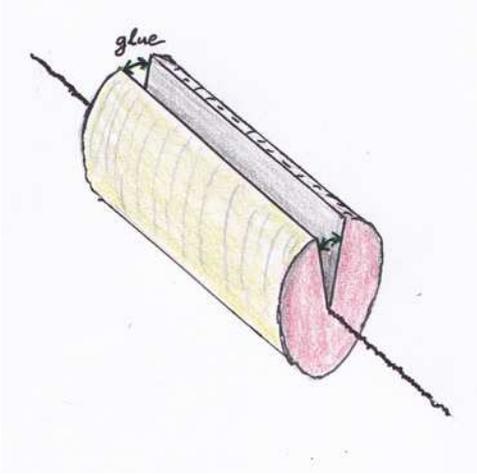}
  \caption{The conical spacetime of a cosmic string}\label{fig-1}
\end{figure}

We will investigate if  chronology protection is still saved in warped 5-dimensional spacetimes.
\begin{figure}
  \includegraphics[height=.1\textheight]{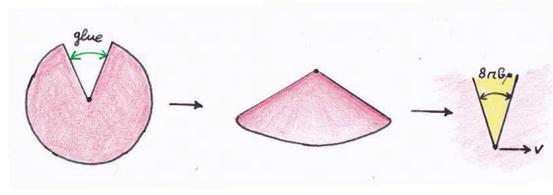}
  \caption{The "cosmon" }\label{fig-2}
\end{figure}
\section{Cosmons: gravitating point particles}\label{cosmon}
\subsection{The 4D case}\label{cosmon a}
It was found several decades ago\cite{garf:1985}, that a self-gravitating U(1)-gauge cosmic string has a conical spacetime at a finite distance of the core. The spacetime becomes
\begin{equation}
ds^2=-e^{a_0}(dt^2-dz^2)+dr^2+e^{-2a_0}(k_2r+a_2)^2d\varphi^2,\label{eq1}
\end{equation}
which can be transformed to Minkowski minus a wedge
\begin{equation}
ds^2=-dt^2+dz^2+dr^2+r^2d\varphi '^2,\label{eq2}
\end{equation}
where now $0<\varphi ' <2\pi e^{-a_0}k_2$. The angle deficit becomes $\Delta\theta = 2\pi (1-e^{-a_0}k_2 )$. One usually write the asymptotic spacetime of Eq.(1) as
\begin{equation}
ds^2=-dt^2+dz^2+dr^2+(1-4 G \mu )^2 r^2 d\varphi^2,\label{eq3}
\end{equation}
\begin{figure}
  \includegraphics[height=.25\textheight]{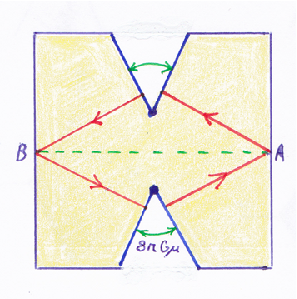}
  \caption{The Gott 2-particle spacetime}
\end{figure}
where $\mu$ represents the mass per unit length of the string and G Newton's constant. So $\Delta\theta = 8\pi G\mu$.
One must realize that this specific solution depends crucial on the energy-momentum tensor of the scalar-gauge field. Some authors put in the line singularity ad hoc by defining $T^{00}=m\delta^2(r)$. Then it is rather trivial that a conical spacetime emerges\cite{star:1963}.
If one omits the z-coordinate, one obtains a gravitating point particle, with a geometry which is locally flat. The particle spacetime is defined via the rotation generator $\Omega$ and Lorentz transformation L,
\begin{equation}
\Omega=\left(
           \begin{array}{ccc}
             \cos 2\alpha & \sin 2\alpha&0 \\
             -\sin 2\alpha & \cos 2\alpha&0 \\
             0&0&1 \\
           \end{array}
         \right)
         L=\left(
           \begin{array}{ccc}
             \cosh\xi & 0&\sinh\xi  \\
             0 & 1&0 \\
             \sinh\xi&0&\cosh\xi \\
           \end{array}
         \right),\label{eq4}
\end{equation}
with $\alpha = 4\pi G\mu$ and $\tanh\xi = v$. A moving particle is obtained by boosting to the rest frame of the particle, rotating and then boosting back.
The generator becomes $T=L\Omega L^{-1}$.
When we construct is this way a two particle generator (with opposite boosts and equal mass) we obtain $T_{eff}=T_1 T_2 =L_1\Omega L_1^{-1} L_2\Omega L_2^{-1}=L_{eff}\Omega_{eff}L_{eff}^{-1}$. This will correspond to a single effective physical source if $\Omega_{eff}$ is a pure rotation through $8\pi G \mu_{eff}$ and $L_{eff}$ a boost ( we can always choose the overall Lorentz frame so that  we may take $T_{eff} = \Omega_{eff}$).
When we compute the traces, we obtain respectively ${\bf Tr}[T_{eff}]=1+2\cos 2\alpha_{eff}=4\cos^2\alpha_{eff} -1$ and ${\bf Tr}[T_1 T_2]=(4\cosh^2\xi\sin^2\alpha -2)^2-1$. From the condition $\cos\alpha_{eff} <1$ we then obtain $\cosh\xi \sin\alpha <1$, or $v<\cos\alpha$. For a Gott spacetime, where a CTC would emerge, the opposite is required\cite{gott:1990}.
That means that the effective mass is imaginary and the identification is boostlike rather than rotationlike. Such a boost-identified spacetime will never arise by boosting a rotation-identified spacetime. This can be easily  seen by the following\cite{deser2:1992}: the global properties of the exterior flat spacetime are characterized  by the identification of points according to $x'=\Omega_{eff} x+c=a_1+L_1\Omega_1 L_1^{-1}\Bigl[a_2-a_1+L_2\Omega_2 L_2^{-1}(x-a_2)\Bigr]$, where $a_i$ are the locations of the particles and $c$ a translation vector. It is easy to see\cite{deser1:1983} that the time-component of $c$ is $2\alpha_{eff}(\vec{a_1}-\vec{a_2})\otimes \vec{v}$, just the orbital angular moment ${\cal J}$ of the effective particle at rest in the c.o.m.(we place the c.o.m. particle in the origin $c^i$=0). But that means that the spacetime around the effective particle is of the form (intrinsic spinning stationary cosmon)
\begin{equation}
ds^2=-(dt+4G{\cal J} d\varphi )^2+dr^2+(1-4G\mu)^2 r^2d\varphi^2.\label{eq5}
\end{equation}
The observables in this model are thus ${\bf Tr}T_{eff}$ and ${\cal J}$. If one transforms this metric to Minkowski minus a wedge, then we have a helical structure of time: when $\varphi$ reaches $2\pi$, t jumps by $8\pi G{\cal J}$. This metric has a singularity because
$T^{00}\sim 4G\mu\delta^2(r), T^{0i}\sim{\cal J}\epsilon ^{ij}\partial_j\delta^2(r)$, describing a spinning point source. There is a CTC if for small enough region ${\cal J}>(1-4G\mu)r_0/4G$.
For a single cosmon, we need not worry about this local CTC. This is manifest, as any classical spin involving derivatives of delta-functions. But we are dealing here with two spinless structureless moving cosmons which gathered  orbital angular momentum. It turns out, as we saw above, that in this case  the CTC criterium will not be satisfied, because the identification is not boostlike as needed in Gott's construction. If the identification would be boostlike, then the conical  structure of the (x,y)-plane is replaced by one in the (t,v)-plane together with a jump in the spatial dimension. Gott's CTC comes towards the interaction region from spaclike infinity. Such a boundary condition is unphysical. In a closed spacetime, one can proof\cite{hooft3:1993,welling:1997} that spacetime shrinks to a point ( big crunch) before the CTC emerges.
\begin{figure}
  \includegraphics[height=.28\textheight]{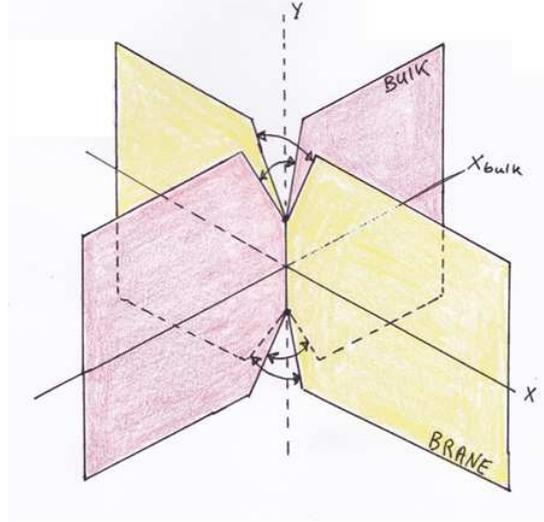}
  \caption{Double conical brane-bulk spacetime}
\end{figure}
\subsection{The 5D case}\label{cosmon a}
Let us now investigate the cosmons in a warped 5D  spacetime, as pictured in figure 4. We again try to find a rotationlike two particle generator via the identification ( we used the coordinates sequence $(y,x,x_{bulk},t)$)
\begin{equation}
T_{eff}=\Bigl(L_1\Omega_{x}\Omega_{xx_B}\Omega_{x_B}L_1^{-1}\Bigr)\Bigl(L_2\Omega_{x}\Omega_{xx_B}\Omega_{x_B}L_2^{-1}\Bigr),\label{eq6}
\end{equation}
with
\begin{eqnarray}
\Omega_x=\left(
           \begin{array}{cccc}
             \cos 2\alpha_1 & \sin 2\alpha_1 &0&0 \\
             -\sin 2\alpha_1 & \cos 2\alpha_1 &0&0 \\
             0&0&1&0 \\
             0&0&0&1 \\
           \end{array}
         \right),\cr
\Omega_{x_B}=\left(
           \begin{array}{cccc}
             \cos 2F\alpha_2 &0& \sin 2F\alpha_2 &0 \\
             0&1&0&0 \\
             -\sin 2F\alpha_2 &0& \cos 2F\alpha_2 &0 \\
             0&0&0&1 \\
           \end{array}
         \right),\label{eq7}
\end{eqnarray}
$L_i$ Lorentz boosts in opposite  x-direction, F the warp factor,
\begin{eqnarray}
L=\left(
           \begin{array}{cccc}
              1&0&0&0\\
            0& \cosh \xi & 0 &\sinh \xi \\
             0&0&1&0 \\
            0& \sinh \xi & 0 &\cosh \xi \\
           \end{array}
         \right)\label{eq8}
\end{eqnarray}
and
\begin{eqnarray}
\Omega_{xx_B}=\left(
           \begin{array}{cccc}
             1 & 0&0 &0 \\
             0&\cos \frac{\pi}{2}&\sin \frac{\pi}{2}&0 \\
             0&-\sin \frac{\pi}{2}&\cos \frac{\pi}{2}&0 \\
             0 & 0&0 &1 \\
           \end{array}
         \right).\label{eq9}
\end{eqnarray}
\begin{figure}
  \includegraphics[height=.28\textheight]{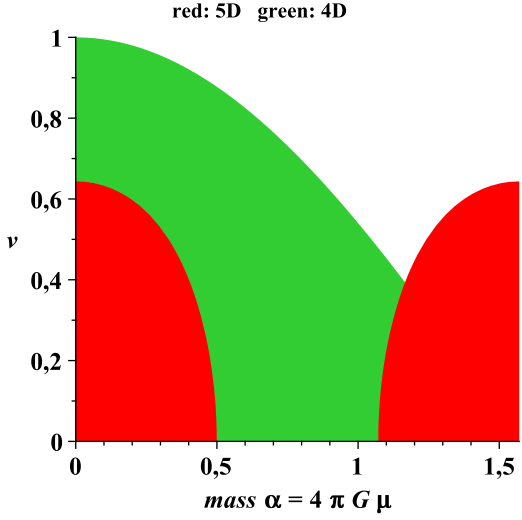}
  \caption{Velocity restriction of a two particle generator in 5D compared with the 4D counterpart situation  $ v<\cos\alpha$}
\end{figure}
\begin{figure}
  \includegraphics[height=.28\textheight]{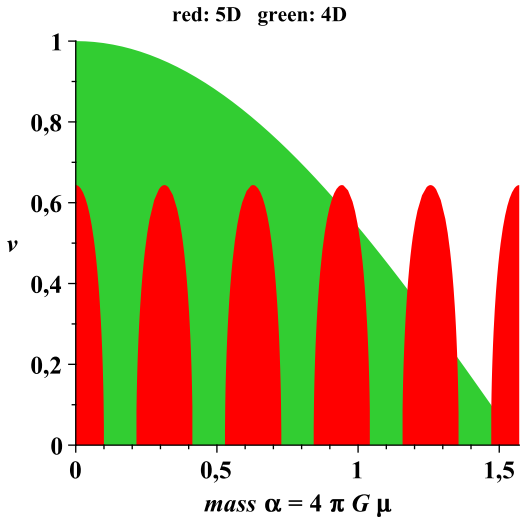}
  \caption{As figure 5, now with warped angle deficit in the bulk, $F=9$ }
\end{figure}
In order to obtain a rotation-identified spacetime, we compare the traces left and right in Eq.(6) and demand that $\cos \alpha_{eff}$ is again $<1$. We obtain the condition (for equal angle deficit in the bulk and brane, $F\alpha_2 =\alpha_1\equiv \alpha$)
\begin{equation}
v<\frac{\sqrt{1+2\sqrt{2}\cos 4\alpha +\cos^2 4\alpha}}{1+\sqrt{2}+\cos 4\alpha}.\label{eq10}
\end{equation}
In figure 5 we plotted this condition and compared it with the 4D condition $v<\cos \alpha $. We see that the rotationlike generator is extended into the Gott region $v> \cos\alpha $.
If we would have taken a rotation in the $(x-x_b)$-plane of $\frac{3}{2}\pi$ in stead of $\frac{1}{2}\pi$, we obtain the maximum value of v=$\sqrt{\sqrt{2}-1}$ in figure 5.
The time-component of the translation vector $c$ becomes  $4\alpha_{eff}(\vec{a_1}-\vec{a_2})\otimes \vec{v}+(a_1-a_2)_xv_x$. The first term is again the angular momentum. The second term is new and disturbs the helical structure in time.

On a warped spacetime, the metric Eq.(3) becomes
\begin{equation}
ds^2 = F(y)\Bigl[-dt^2+dz^2+dr^2+(1-4 G \mu )^2 r^2 d\varphi^2\Bigr]+dy^2,\label{eq11}
\end{equation}
with $F$ the warp factor. One has now a warped mass per unit length $F\mu$  in the bulk. In figure 6 we plotted the velocity restriction for the warpfactor $F=9$.
We can, however, always recover the solution of figure 5 for any tiny bulk-angle deficit by choosing a large enough warp factor.

\section{The Effective Brane}\label{brane}
We previously investigated the effective brane equations of a self-gravitating  U(1)-gauge cosmic string\cite{slagter1:2012} on the warped spacetime
\begin{equation}
ds^2=F(y)\Bigl[e^{A(r)}(-dt^2+dz^2)+dr^2+\frac{K(r)^2}{e^{2A(r)}}d\varphi^2\Bigr]+dy^2,\label{eqn12}
\end{equation}
with $ K$ and $A$ the metric functions.
The changes of the brane equations are trigged by the projection of the 5D Weyl tensor on the brane and a bulk cosmological constant.
The effective Einstein equations are\cite{maart:2003}
\begin{eqnarray}
{^{(4)}G}_{\mu\nu}=-\Lambda_{eff}{^{(4)}g}_{\mu\nu}+\kappa_4^2 {^{(4)}T}_{\mu\nu}+\kappa_5^4{\cal S}_{\mu\nu}\cr -{\cal E}_{\mu\nu}
+\frac{2}{3}\kappa_5^4{\cal F}_{\mu\nu},\label{eq13}
\end{eqnarray}
where $\Lambda_{eff}=\frac{1}{2}(\Lambda_5+\kappa_4^2\lambda_4)=\frac{1}{2}(\Lambda_5+\frac{1}{6}\kappa_5^4\lambda_4^2)$,
 $\lambda_4$ is the vacuum energy in the brane (brane tension), and ${\cal F}_{\mu\nu}$ the energy-stress tensor contribution from the bulk scalar field.
The first correction term ${\cal S}_{\mu\nu}$ is  the quadratic term in the
energy-momentum tensor arising from the extrinsic curvature terms in the projected Einstein tensor. The second correction term ${\cal E_{\mu\nu}}$ is a part of the 5D Weyl tensor and carries information of the gravitational field outside the brane. It turns out that the equations for the scalar-gauge fields, in the empty bulk case, ${\cal F}_{\mu\nu}=0$, do not change with respect to the 4D case. The equations for the metric functions $A$ and $K$ change dramatically. Defining $\Theta_1\equiv K\partial_r A, \Theta_2\equiv \partial_r K$, the equations can be written as
\begin{eqnarray}
\partial_r\Theta_2=\frac{6K}{11}\Bigl[\frac{3c_1}{4}-6\Lambda_{eff}+\kappa_4^2(3\varrho_r-2\sigma +\varrho_\varphi ) \cr +\kappa_5^4(3\xi_r-2\xi_t  +\xi_\varphi )\Bigr],\label{eq14}
\end{eqnarray}
\begin{eqnarray}
4\partial_r\Theta_1+\partial_r\Theta_2=
6K\Bigl[\frac{c_1}{4}-2\Lambda_{eff}+\kappa_4^2(\varrho_r +\varrho_\varphi)\cr  +\kappa_5^4(\xi_r +\xi_\varphi)\Bigr],\label{eq15}
\end{eqnarray}
with $c_1$ a constant and where we write the stress energy tensors as
$T_{\mu\nu}=\sigma \hat{k}_t\hat{k}_t+\varrho_z\hat{k}_z\hat{k}_z+\varrho_\varphi \hat{k}_\varphi\hat{k}_\varphi+\varrho_r\hat{k}_r\hat{k}_r,
 {\cal S}_{\mu\nu}=\xi_t \hat{k}_t\hat{k}_t+\xi_z\hat{k}_z\hat{k}_z+\xi_\varphi \hat{k}_\varphi\hat{k}_\varphi+\xi_r\hat{k}_r\hat{k}_r$,
with $\hat{k}_i$ a set of orthonormal vectors. In the 4D case, they are\cite{garf:1985}
\begin{eqnarray}
\partial_r\Theta_2 =\frac{1}{2}\kappa_4^2K(3\varrho_r -2\sigma +\varrho_\varphi) \cr \partial_r\Theta_1=\kappa_4^2 K(\varrho_r+\varrho_\varphi).\label{eq16}
\end{eqnarray}
For $c_1=8\Lambda_{eff}$, we see on the right hand sides the same combinations of the energy-momentum tensor components. The left hand sides are different. This is a consequence of
the propagation of the gravitons into the bulk, i.e., the appearance of the ${\cal E_{\mu\nu}}$ term. When we use a clever combination of the field equations and the
conservation of stress energy\cite{garf:1985}
\begin{eqnarray}
\hat{k}_\mu\nabla_\nu T^{\nu\mu}= \partial_r(K\varrho_r)-\varrho_\varphi\partial_r K
+(\sigma +\varrho_\varphi)K\partial_r A=0.\label{eq17}
\end{eqnarray}
one then obtains in the 5D case the expression\cite{slagter1:2012}
\begin{equation}
\partial_r(\kappa_4^2K^2\varrho_r)=\partial_r\Bigl[\frac{2}{3}\Theta_1\Theta_2+\frac{1}{12}\Theta_2^2-\frac{1}{2}\Theta_1^2\Bigr].\label{eq18}
\end{equation}
If we assume that $\int_0^\infty K\sigma dr$ converges,  $\sigma >|\varrho_r|$ and that $\lim\limits_{r\rightarrow \infty}K^2\sigma=0$, we obtain from Eq.(18)
\begin{equation}
\bar \Theta_1=\frac{1}{-4\pm \sqrt{22}}\bar \Theta_2,\label{eq19}
\end{equation}
where we denote with $\bar \Theta_1$ and $\bar \Theta_2$ the asymptotic values.
Let us compare this solution with the 4D counterpart equation
\begin{eqnarray}
\partial_r(\kappa_4^2K^2\varrho_r)=\partial_r\Bigl[\Theta_1(\Theta_2-\frac{3}{4}\Theta_1)\Bigr].\label{eq20}
\end{eqnarray}
Then we have two possibilities: $\bar \Theta_1=0$, or $\bar\Theta_2=\frac{3}{4}\bar\Theta_1$, which represents a Kasner-type solution. The first possibility, $\bar \Theta_1=0$, represents the asymptotic spacetime of Eq.(1) and can be transformed to the conical spacetime of Eq.(2), i.e., Minkowski minus a wedge, with angle deficit $\Delta\theta=2\pi(1-e^{-a_0}k_2)$\cite{garf:1985}.

In our model, the asymptotic spacetime becomes
\begin{eqnarray}
ds^2=-e^{a_0}(k_2 r+a_2)^{\frac{1}{-4\pm\sqrt{22}}}(dt^2-dz^2)+dr^2 \cr +e^{-2a_0}(k_2 r+a_2)^{2+\frac{2}{4\mp\sqrt{22}}}d\varphi^2,\label{eq21}
\end{eqnarray}
which is not conical!
\section{Conclusions}\label{concl}
We calculated the matching conditions of a Gott pair of particles in a warped spacetime. It turns out that the center of mass of the effective particle is not tachyonic and the Gott condition could be fulfilled because there is a velocity overlap. The time component of the translation vector has a term still proportional to the orbital angular momentum and an extra term, which spoils the time-helical structure. Further, it turns out that the effective brane field equations of a gravitating U(1) gauge field, the basis of the conical structure outside the cosmon, suffers  a conical feature.

\end{document}